\documentclass[9pt,twocolumn,twoside]{opticajnl}
\journal{opticajournal} 

\setboolean{shortarticle}{true}

\usepackage[dvipsnames]{xcolor}
\usepackage[acronym]{glossaries}
\usepackage[textwidth=1.6cm]{todonotes}
\usepackage{comment}
\usepackage{nameref}
\usepackage{subcaption}
\usepackage{multicol}
\usepackage{multirow}
\usepackage{rotating}
\usepackage{stfloats}
\usepackage[noadjust]{cite}
\usepackage{float}
\usepackage{amsfonts}
\usepackage{upgreek}
\usepackage{mathtools}
\usepackage{scalerel}
\usepackage{braket}
\usepackage{tcolorbox}
\usepackage{lineno}
\usepackage{lipsum}
\usepackage{siunitx}

\newacronym{sdm}{SDM}{space-division multiplexing}
\newacronym{mmf}{MMF}{multi-mode fiber}
\newacronym{si-mmf}{SI-MMF}{step-index multi-mode fiber}
\newacronym{gi-mmf}{GI-MMF}{graded-index multi-mode fiber}
\newacronym{smf}{SMF}{single-mode fiber}
\newacronym{mcf}{MCF}{multi-core fiber}
\newacronym{osa}{OSA}{optical spectral analyzer}
\newacronym{uc-mcf}{UC-MCF}{uncoupled-core multi-core fiber}
\newacronym{cc-mcf}{CC-MCF}{coupled-core multi-core fiber}
\newacronym{srs}{SRS}{stimulated Raman scattering}
\newacronym{sprs}{SpRS}{spontaneous Raman scattering}
\newacronym{qkd}{QKD}{quantum key distribution}
\newacronym{ase}{ASE}{amplified spontaneous emission}
\newacronym{otdr}{OTDR}{optical time-domain reflectometer}
\newacronym{psd}{PSD}{power spectral density}

\newcommand{\der}{\mathrm{d}}
\newcommand{\diag}{\mathrm{dg}}
\newcommand{\R}{\mathrm{R}}
\newcommand{\Total}{\mathrm{T}}
\newcommand{\Th}{\mathrm{th}}
\newcommand{\REF}{\mathrm{ref}}

\title{Spontaneous Raman scattering in SDM fibers}

\author[1,*]{Lucas~Alves~Zischler}
\author[1]{Giammarco~Di~Sciullo}
\author[1]{Divya~A.~Shaji}
\author[1]{Antonio~Mecozzi}
\author[1]{Cristian~Antonelli}

\affil[1]{Department of Physical and Chemical Sciences, University of L’Aquila, 67100 L’Aquila, Italy}
\affil[*]{lucas.zischler@univaq.it}

\begin{abstract}
  \Gls*{sprs} is a weak non-linear effect, particularly relevant to classical-quantum coexistence transmission and sensing applications. In classical transmission, the relevant Raman effect is \gls*{srs}, and recent studies have examined it in \gls*{sdm} fibers. An intrinsic relation between \gls*{sprs} and \gls*{srs} allows previous \gls*{srs} results to inform \gls*{sprs} models. In this work, we extend \gls*{sprs} models derived for \glspl*{smf} to \gls*{sdm} fibers with multiple mode groups of degenerate modes, covering both Stokes and anti-Stokes bands. The proposed model is a useful, fiber-design-independent tool for evaluating scattered noise in optical links, and it is validated through experimental measurements in field-deployed \glspl*{mcf} and \gls*{mmf}, showing good agreement.
\end{abstract}

\setboolean{displaycopyright}{false}

\begin{document}

\maketitle

\glsresetall

In recent years, \gls*{sdm} has emerged as one of the leading technologies to sustain increasing traffic demands~\citep{winzer2012optical}, and the additional spatial dimension has been exploited to benefit \gls*{qkd} transmission~\citep{eriksson2019inter}. Nevertheless, there is a lack of studies on \gls*{sprs} in \gls*{sdm} fibers, a phenomenon particularly crucial in classical-quantum coexistence scenarios~\citep{kumar2015coexistence}.

In classical transmission, the relevant Raman effect is \gls*{srs}, and recent work have evaluated it in \glspl*{mcf}~\citep{di2025characterization} and \glspl*{mmf}~\citep{schneck2025experimental,zischler2025evaluation,divya2026raman}. Since both \gls*{srs} and \gls*{sprs} arise from the delayed Kerr response of silica, a simple relation linking the two efficiencies can be expected. This work provides analytical formulas linking \gls*{sprs} noise power to Raman gain efficiency. We extend previously established \gls*{sprs} models developed for the Stokes region in \glspl*{smf}~\citep{rottwitt2003scaling,bromage2004raman} to \gls*{sdm} fibers, including the anti-Stokes region. The proposed model addresses the most general case of SDM fibers featuring multiple groups of degenerate modes. The model is experimentally validated with measurements performed in a 4-core \gls*{uc-mcf}, 4-core \gls*{cc-mcf}, and a 15-mode graded-index \gls*{mmf} field-deployed in the city of L'Aquila, Italy~\citep{antonelli2025space}. Results show good agreement between theory and measurements.

Raman scattering can be described in quantum physics as a process where a pumped optical phonon decays to a lower energy state, with the Stokes shift determined by molecular vibrational losses. In classical wave theory, it corresponds to the delayed third-order non-linear susceptibility of the medium.

In the scalar case, neglecting losses of the fiber medium, the mean rate of scattered photons $N_{s}$ in an infinitesimal segment $\der z$, guided along the fiber with energy $hf_{s}$ (where $h$ is Planck's constant and $f_{s}$ the photon frequency), is given by~\citep[Eq.~(6)]{bromage2004raman},~\citep[Eq.~(11)]{rottwitt2003scaling},~\citep[Eq.~(21)]{lin2007photon}.
\begin{equation}
  \pm\frac{\der N_{s}(z)}{\der z}=\begin{cases}
    [1+\Phi(\Delta f)]g_{\R}(f_{p},\Delta f)P_{p}(z), & f_{s}<f_{p},\\
    \Phi(\Delta f)g_{\R}(f_{p},\Delta f)P_{p}(z), & f_{s}>f_{p},
  \end{cases}
  \label{eq:dndz}
\end{equation}
where, $\pm$ accounts for the propagation direction, $f_{p}$ is the pump frequency, and $\Phi(\Delta f)$ is the thermally populated phonon occupancy factor, discussed in more detail below, with ${\Delta f = |f_{s} - f_{p}|}$. The coefficient $g_{\R}$ represents the \gls*{srs} efficiency, and $P_{p}(z)$ denotes the optical power at the pump frequency $f_{p}$ and distance $z$. The \gls*{srs} coefficient accounts only for the portion of the Raman cross-section resulting in guided photons. The photon rate can be expressed in power units as ${P^{\mathrm{SpRS}}_{s}(z) = hf_{s} B_{\REF} N_{s}(z)}$, where $B_{\REF}$ is a reference bandwidth. In contrast to \gls*{srs}, \gls*{sprs} produces scattered photons in both propagation directions.

In the Stokes region, the factor $[1+\Phi(\Delta f)]$ accounts for phonons in the ground vibrational state plus thermally populated phonons with vibrational energy $h \Delta f$, which follow Bose-Einstein statistics ${\Phi(\Delta f)=\left\{\exp\left[h\Delta f/(k_{B}T)\right]-1\right\}^{-1}}$.

In the anti-Stokes process, only phonons in a higher vibrational state produce scattered light, when decaying to lower energy states. Albeit weaker, anti-Stokes \gls*{sprs} is always present in real setups, as thermally loaded phonons can provide the energy needed to produce photons with a positive frequency shift.

When multiple spatial modes and polarizations are present, the Raman effect can encompass all distinct spatial paths. The \gls*{srs} efficiency from the $h^{\Th}$ to the $i^{\Th}$ mode is given by~\citep{antonelli2013raman}
\begin{equation}
  g_{\R,i,h}(f_{p},\Delta f)=\frac{\pi}{4}f_{s}\chi^{(3)}\epsilon_{0}h_{\R}(\Delta f)A^{-1}_{\text{eff},i,h}(f_{p},f_{s}),
  \label{eq:srseff}
\end{equation}
where $\chi^{(3)}$ is the third-order susceptibility coefficient, $h_{\R}(\Delta f)$ is the material's Raman parallel-polarized impulse response\footnote{The cross-polarized response is neglected due to its negligible magnitude in silica fibers.}~\citep[Eq.~(6)]{hellwarth1975origin}, and $\epsilon_{0}$ is the vacuum permittivity. We approximate the fiber as a homogeneous medium, so the Raman impulse response is independent of the mode profile, where this assumption is later validated with experimental measurements. The self- or cross-effective mode area ${A_{\text{eff},i,h}(f_{p},f_{s})}$ is given by~\citep[Eq.~(7)]{poletti2008description}
\begin{equation}
  A_{\text{eff},i,h}(f_{s},f_{p})\hspace{-3pt}=\hspace{-3pt}\frac{\iint\hspace{-2pt}||\vec{\mathbf{F}}_{i}(x,y,f_{s})||^{2}\der x\der y\hspace{-2pt}\iint\hspace{-2pt}||\vec{\mathbf{F}}_{h}(x,y,f_{p})||^{2}\der x\der y}{\iint||\vec{\mathbf{F}}_{i}(x,y,f_{s})\cdot\vec{\mathbf{F}}_{h}(x,y,f_{p})||^{2}\der x\der y}\hspace{-1pt},
\end{equation}
where $\vec{\mathbf{F}}_{i}(x,y,f)$ is the $i^{\Th}$ mode profile at frequency $f$.

It is therefore convenient to define a mode-independent Raman response that can be extracted from the \gls*{srs} efficiency, defined as
\begin{equation}
  \hat{g}_{\R}(f_{p},\Delta f)=\frac{\pi}{4}f_{s}\chi^{(3)}\epsilon_{0}h_{\R}(f_{p},f_{s}).
  \label{eq:grnorm}
\end{equation}

The modes can be grouped into weakly-coupled groups of degenerate modes. The \gls*{srs} gain efficiency, averaged over the mode groups, scales by the inverse of the self- and cross-mode-group-averaged effective areas, given by
\begin{equation}
  \tilde{A}_{\text{eff},n,m}(f_{s},f_{p})=\frac{4N_{n}(f_{s})N_{m}(f_{p})}{\sum_{\substack{i\in[1,2\cdot N_{n}(f_{s})]\\h\in[1,2\cdot N_{m}(f_{p})]}}A^{-1}_{\text{eff},i,h}(f_{p},f_{s})},
  \label{eq:aeffmg}
\end{equation}
where $N_{n}(f)$ is the number of spatial modes in the $n^{\Th}$ mode group at frequency $f$. The indices $i$ and $h$ span all spatial modes and polarizations. The mode-group-averaged \gls*{srs} efficiency is then given by~${\tilde{g}_{\R,n,m}(f_{p},\Delta f)=\hat{g}_{\R}(f_{p},\Delta f)\tilde{A}^{-1}_{\text{eff},n,m}(f_{s},f_{p})}$.

Under the assumption that all modes within a given mode group $n$ are statistically equivalent, the mode-group-averaged \gls*{sprs} power satisfies the following evolution equation
\begin{equation}
  \begin{aligned}
    \pm\frac{\der P^{\mathrm{SpRS}}_{s,n}(z)}{\der z}=&-\alpha_{n}(f_{s})P^{\mathrm{SpRS}}_{s,n}(z)+\sum_{m\neq n}\kappa_{n,m}(f_{s})P^{\mathrm{SpRS}}_{s,m}(z)\\
    &+\sum_{m}\eta_{n,m}(f_{p},\Delta f)P^{\Total}_{p,m}(z),
  \end{aligned}
  \label{eq:dpdz}
\end{equation}
where $\alpha_{n}(f)$ is the mode-group-averaged attenuation profile, $\kappa_{n,m}(f)$ the coupling coefficient between mode groups $n$ and $m$, and $P^{\Total}_{p,m}(z)$ is the total power in mode group $m$ at frequency $f_{p}$ and distance $z$. The coefficient $\eta_{n,m}(f_{p},\Delta f)$ represents the \gls*{sprs} efficiency from the $m^{\Th}$ to the $n^{\Th}$ mode group, given by
\begin{equation}
  \eta_{n,m}(f_{p},\Delta f)=\begin{cases}
    [1+\Phi(\Delta f)]hf_{s}B_{\REF}\tilde{g}_{\R,n,m}(f_{p},\Delta f), & f_{s}<f_{p},\\
    \Phi(\Delta f)hf_{s}B_{\REF}\tilde{g}_{\R,n,m}(f_{p},\Delta f), & f_{s}>f_{p}.
  \end{cases}
  \label{eq:etagr}
\end{equation}

Similarly to the \gls*{srs} efficiency, a \gls*{sprs} coefficient $\hat{\eta}(f_{p},\Delta f)$ solely dependent on the material characteristics can be defined as
\begin{equation}
  \eta_{n,m}(f_{p},\Delta f)=\hat{\eta}(f_{p},\Delta f)\tilde{A}^{-1}_{\text{eff},n,m}(f_{s},f_{p}).
  \label{eq:etamat}
\end{equation}

We utilized the proposed model to characterize the \gls*{sprs} efficiency on two types of \glspl*{mcf}, \gls*{uc-mcf} and \gls*{cc-mcf}, both with 4 cores and equal single-core self-effective mode area (\SI{81}{\um\squared} at \SI{1550}{\nm}~\citep{hayashi2019field}, \SI{162}{\um\squared} polarization-mode averaged cross-effective area). The cores of the \gls*{uc-mcf} are single-mode and experience minimal inter-core coupling. In the \gls*{cc-mcf}, the cores are strongly coupled and the local modes form a single group of degenerate modes.

All experimental measurements were performed with a Raman pump at \SI{1455}{\nano\meter}. The \gls*{srs} measurements were performed in \citep{di2025characterization}, and are briefly reviewed here. The \gls*{srs} efficiency was estimated from the on-off Raman gain of a probe signal, which was swept from \SI{1500}{\nano\meter} to \SI{1620}{\nano\meter}. The probe was co- and counter-propagated relative to the pump in the \gls*{cc-mcf}. The signal power was measured with an \gls*{osa} both with the pump on and off. To remove the SpRS contribution from the on-off Raman gain measurements, an additional measurement was performed with the pump on and the signal off. At the fiber input, the pump was attached to a single core in both fiber types. The optical power was measured in the core under test in the \gls*{uc-mcf}, and simultaneously measured from all fiber cores with a power coupler in the \gls*{cc-mcf} case.

In the \gls*{uc-mcf} measurements, the pump was launched in a 25.3-\si{\kilo\meter} long fiber link with measured optical power of \SI{17.95}{\decibel m}. For the \gls*{cc-mcf} measurements, the pump was launched in a 69.3-\si{\kilo\meter} long link with measured optical power of \SI{29}{\decibel m}. In both scenarios, the probe signal was launched with a measured optical power of \SI{-35}{\decibel m} per core.

\begin{figure*}[!t]
    \centering
    \includegraphics{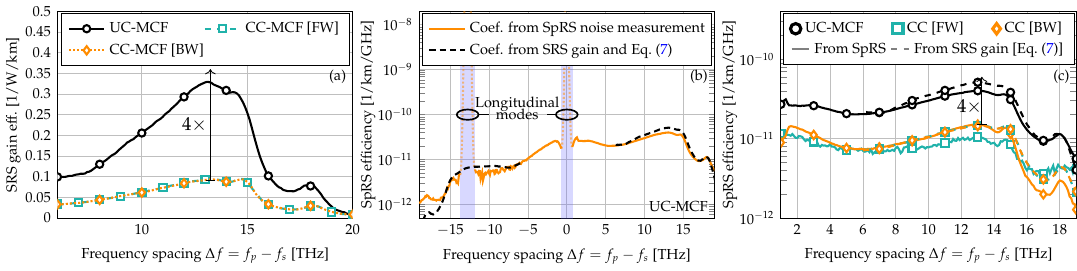}
    \caption{\justifying (a) Mode–group–averaged \gls*{srs} gain efficiency measured in the \gls*{uc-mcf} (solid line, circles), copropagating \gls*{cc-mcf} (dashed, squares), and counter-propagating \gls*{cc-mcf} (dotted, diamonds) configurations. Gain curves for the \gls*{cc-mcf} are extracted from~\citep{di2025characterization}. (b) Estimated \gls*{sprs} efficiency in the 4-core \gls*{uc-mcf}, obtained from copropagating noise (solid lines) and on–off gain (dashed lines) data, normalized by bandwidth. Shaded regions indicate the bandwidth of the unfiltered longitudinal laser modes. (c) Estimated \gls*{sprs} efficiency from experimental measurements in the 4-core \gls*{uc-mcf} (circles) and 4-core \gls*{cc-mcf} (squares: copropagating, diamonds: counter-propagating), obtained from noise (solid lines) and on-off gain (dashed lines) data. The labels ``\textit{FW}'' and ``\textit{BW}'' represent, respectively, co- and counter-propagating measurements.}
    \label{fig:Mcf}
\end{figure*}

The results for the mode-group-averaged \gls*{srs} efficiency are shown in Fig.~\ref{fig:Mcf}(a). In the \gls*{cc-mcf}, the optical power is distributed over an effective cross-section area four times larger, reducing the group-averaged \gls*{srs} coefficient by a factor of four.

For the \gls*{cc-mcf} and for each core of the \gls*{uc-mcf}~\eqref{eq:dpdz} reduces to differential equations of the same form. Solving the reduced equation for $P^{\mathrm{SpRS}}_{s,n}(z)$ yields an analytical relation between group-averaged \gls*{sprs} efficiency and fiber parameters, given as
\begin{equation}
  \eta(f_{p},\Delta f) = \begin{cases}
    \frac{P^{\mathrm{SpRS}}_{s}(L)\exp\left[\alpha(f_{s}) L\right]}{\int_{0}^{L} P^{\Total}_{p}(z')\exp\left[\alpha(f_{s}) z'\right]dz'}\hspace{-1pt}, &\text{Co-prop.},\\
    \frac{P^{\mathrm{SpRS}}_{s}(0)}{\int_{0}^{L} P^{\Total}_{p}(z')\exp\left[-\alpha(f_{s}) z'\right]dz'}\hspace{-1pt}, &\text{Counter-prop.},
  \end{cases}
  \label{eq:sprsscalar}
\end{equation}
where $L$ is the fiber length.

For the experimental \gls*{sprs} noise measurements, only the pump wave was transmitted. The pump power at the fiber input was set to \SI{2.7}{\decibel m} in the \gls*{uc-mcf} setup, and maintained at 29~dBm in the \gls*{cc-mcf}. All other experimental settings were identical to those used for the \gls*{srs} gain measurements.

Figure~\ref{fig:Mcf}(b) shows the \gls*{sprs} efficiency for the \gls*{uc-mcf}, where solid lines represent \gls*{sprs} efficiency derived from noise measurements, while dashed lines are the values estimated from the \gls*{srs} gain efficiency using the analytical relation in~\eqref{eq:etagr}, assuming a waveguide temperature of \SI{283}{\kelvin}. Noise measurements are corrupted by the unfiltered laser’s longitudinal modes. In order to remove the contribution of the unfiltered laser's modes and isolate the pump \gls*{sprs} noise, we subtract a scaled version of the measured noise profile, adjusted for each mode launch power and attenuation. The \gls*{sprs} efficiency extracted from noise measurements is in good agreement with the values obtained from the \gls*{srs} gain estimates, with a divergence at the largest anti-Stokes shift due to the device noise floor of the \gls*{osa}.

In Fig.~\ref{fig:Mcf}(c), we plot the \gls*{sprs} efficiency for the \gls*{uc-mcf} and \gls*{cc-mcf}. For the \gls*{cc-mcf}, the \gls*{sprs} efficiency derived from \gls*{srs} gain is identical for the co- and counter-propagating cases, as the Raman gain is independent of signal direction~\citep{di2025characterization}. Curves from noise measurements, however, show small differences between both directions, although \gls*{sprs} is theoretically direction-independent. Overall, the \gls*{sprs} spectra for both fibers in all scenarios match in shape and are close in magnitude to those obtained from the \gls*{srs} gain efficiency values.

As previously observed, in Fig.~\ref{fig:Mcf}(c), the \gls*{cc-mcf} curves show a four-fold reduction in the group-averaged \gls*{sprs} efficiency compared to the \gls*{uc-mcf}, due to the larger number of degenerate modes. Therefore, for the same pump power per core, each core of the \gls*{uc-mcf} accumulates the same \gls*{sprs} noise as in the \gls*{cc-mcf}.

\begin{figure}[!b]
    \centering
    \includegraphics{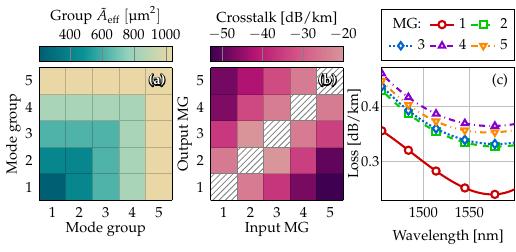}
    \caption{\justifying (a) Self- and cross-effective areas of \gls*{mmf} mode-groups. (c) Mode-group-averaged coupling coefficients and (c) mode-group-averaged frequency-dependent attenuation of the \gls*{mmf}.}
    \label{fig:MmfParam}
\end{figure}

The same characterization was performed on a 15-mode (5-mode-group) 50-\si{\kilo\meter} long graded-index \gls*{mmf}~\citep{sillard2016low}, deployed in the same testbed. The mode-group-averaged effective and cross-effective areas are shown in Fig.~\ref{fig:MmfParam}(a). Within conventional transmission bands, the self- and cross-effective mode areas have a negligible dependence on frequency, and are assumed as constant.

For the \gls*{sprs} characterization, the pump was launched sequentially into each available mode, and the corresponding backscattered noise was measured with the \gls*{osa} for each individual mode. The pump power launched into the mode-multiplexer was measured to be, on average, \SI{19.70}{\decibel m}, with small variations across the selected pump configurations due to differing insertion losses among the optical paths. The crosstalk and mode-dependent losses of the multiplexer were derived from C-band measurements and presented in~\citep{gatto2024partial}.

\begin{figure*}[!t]
    \centering
    \includegraphics{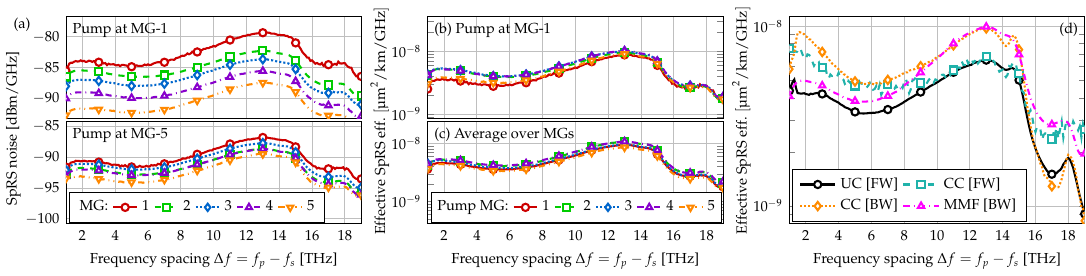}
    \caption{\justifying (a) Per-mode counter-propagating \gls*{sprs} noise in each mode group (distinguished by line styles) of the \gls*{mmf}, with the pump launched into the $1^{\mathrm{st}}$ and $5^{\Th}$ mode groups. (b) Effective \gls*{sprs} efficiency estimated from measurements of each mode group with the pump in the fundamental mode. (c) Mode-group-averaged measurements for each pump configuration. (d) Effective \gls*{sprs} efficiency for the different \gls*{sdm} fiber types, where the \gls*{mmf} curve correspond to the average over all measurements.}
    \label{fig:Mmf}
\end{figure*}

Linear mode coupling must be considered in \glspl*{mmf} measurements, as crosstalk is usually non-negligible. Coupling coefficients are measured for the pump frequency and for the signal band with a 1520-to-\SI{1580}{\nano\meter} noise source. The mode-group- and frequency-averaged values are given in Fig.~\ref{fig:MmfParam}(b). Minor frequency-dependent deviations were observed between coupling coefficients, therefore, the same coupling matrix was considered for the pump and signal. Losses showed significant frequency-dependence, with the mode-group-averaged values plotted in Fig.~\ref{fig:MmfParam}(c), estimated from a polynomial interpolation of the experimentally measured values.

Supplementing~\eqref{eq:sprsscalar} with terms accounting for linear coupling and inter-modal scattering, the material \gls*{sprs} efficiency $\hat{\eta}$ can be estimated from \gls*{sprs} noise values measured in any arbitrary mode-group $n$, as given by
\begin{equation}
  \hat{\eta}(f_{p},\Delta f) = \begin{cases}
    \frac{P^{\mathrm{SpRS}}_{s,n}(L)}{\left\{\int_{0}^{L}e^{[\diag(\boldsymbol{\alpha}_{s})-\mathbf{K}_{s}](z'-L)}\mathbf{A}^{\text{inv}}_{\text{eff}}\vec{P}^{\Total}_{p}(z')dz'\right\}_{n}}\hspace{-1pt}, &\text{Co-prop.},\\
    \frac{P^{\mathrm{SpRS}}_{s,n}(0)}{\left\{\int_{0}^{L}e^{-[\diag(\boldsymbol{\alpha}_{s})+\mathbf{K}_{s}]z'}\mathbf{A}^{\text{inv}}_{\text{eff}}\vec{P}^{\Total}_{p}(z')dz'\right\}_{n}}\hspace{-1pt}, &\text{Counter-prop.},
  \end{cases}
  \label{eq:etaeffmmf}
\end{equation}
where $\mathbf{A}^{\text{inv}}_{\text{eff}}$ is a matrix whose elements are the mode-group-averaged inverse effective areas $\tilde{A}^{-1}_{\text{eff},i,h}$, the function $\diag(\cdot)$ denotes the matrix diagonal operator, $\boldsymbol{\alpha}_{s}$ and $\mathbf{K}_{s}$ are for the signal frequency the attenuation vector and coupling matrix respectively, and $\vec{P}^{\Total}_{p}(z)$ is the total pump power vector at distance~$z$.

The measured mode-group-averaged backscattered \gls*{sprs} noise, is shown in Fig.~\ref{fig:Mmf}(a) for the pump launched in the $1^{\mathrm{st}}$ and $5^{\Th}$ mode groups. We observe significant deviations in the noise levels when the pump was launched into the fundamental mode group, which are notably reduced when the pump was launched into the highest-order mode group. These deviations in noise levels are directly related to the overlap between mode groups, as indicated by the effective-area matrix shown in Fig.~\ref{fig:MmfParam}(a).

From~\eqref{eq:etaeffmmf}, with the pump allocated in the fundamental mode-group, we obtain the \gls*{sprs} efficiency curves shown in Fig.~\ref{fig:Mmf}(b). Estimates derived from noise values across all mode groups are in excellent agreement, with minor deviations near the pump frequency.

In Fig.~\ref{fig:Mmf}(c), we plot the mode-group-averaged measurements for each pump configuration. All measurements are in excellent agreement throughout the entire evaluated region, indicating that the \gls*{sprs} efficiency can be reliably estimated using any pump configuration. Estimates derived from the \gls*{srs} efficiency are not presented, as although Raman gain profiles were measured in the same fiber~\citep{divya2026raman}, measurements of the \gls*{srs} coefficients in \glspl*{mmf} remain an ongoing effort, and no such characterization has yet been reported.

As the material \gls*{sprs} efficiency is independent of fiber design, it can be directly compared across different fiber types. Figure~\ref{fig:Mmf}(d) shows the material \gls*{sprs} efficiency extracted from noise measurements for the \gls*{uc-mcf}, \gls*{cc-mcf}, and \gls*{mmf}. The efficiencies lie within a similar range, showing close agreement between all fiber designs, with the co-propagating \gls*{uc-mcf} estimation marginally lower. While small deviations are expected due to variations in doping concentrations among different fiber designs, direction-dependent discrepancies may arise from backscattering at splices or span connections. Overall, the agreement observed across distinct fiber types indicates that a mode-independent \gls*{sprs} profile can be extracted irrespective of the fiber design, with the mode profiles accounted for solely through the effective-area matrix.

To conclude, we develop a generalized model for the longitudinal evolution of \gls*{sprs} power in \gls*{sdm} fibers with multiple mode groups, extending prior formulations for \glspl*{smf} by decomposing the \gls*{sprs} efficiency as the product of a material-dependent term and a dimensional overlap factor between the interacting modes. Additionally, the model relates the \gls*{sprs} coefficient in both the Stokes and anti-Stokes regions to the \gls*{srs} efficiency. The model is utilized to characterize the \gls*{sprs} efficiency experimentally in three field-deployed \gls*{sdm} fibers, a 4-core \gls*{uc-mcf}, a 4-core \gls*{cc-mcf}, and a 15-mode graded-index \gls*{mmf}. Measurements show good agreement between analytical estimates derived from previously reported \gls*{srs} efficiency data and \gls*{sprs} efficiency derived from the received scattered noise, as well as consistent across the considered fibers. These results advance the understanding of non-linear impairments in \gls*{sdm} links where \gls*{sprs} is non-negligible, a regime that is particularly relevant for the coexistence of classical optical communication and \gls*{qkd}.

\begin{backmatter}
\bmsection{Funding} Funded by the European Union (Grant Agreement 101120422 and 101072409). Views and opinions expressed are, however, those of the author(s) only and do not necessarily reflect those of the European Union or REA. Neither the European Union nor the granting authority can be held responsible for them.

\bmsection{Disclosures} The authors declare no conflicts of interest.

\bmsection{Data availability} Data underlying the results presented in this paper are not publicly available at this time but may be obtained from the authors upon reasonable request.
\end{backmatter}

\bibliography{sample}
\bibliographyfullrefs{sample}

\end{document}